\documentclass[conference]{IEEEtran}
\IEEEoverridecommandlockouts
\usepackage{amsmath}
\usepackage{color}\usepackage{bbm}
\usepackage{multirow}
\usepackage{amsmath}
\usepackage{mwe}
\usepackage{amsthm} 
\usepackage{amssymb}
\usepackage{float}
\usepackage{xurl}
\usepackage{nccmath}
\usepackage{graphicx}
\usepackage{subcaption}
\usepackage[linesnumbered,ruled]{algorithm2e}
\usepackage{booktabs}
\usepackage{mwe}
\usepackage{cite}
\usepackage{graphicx}
\usepackage{textcomp}
\usepackage[normalem]{ulem}
\usepackage{xcolor}
\usepackage{tabulary}
\usepackage{array} 
\usepackage{mathtools}

\usepackage{hyperref}
\definecolor{linkcolour}{rgb}{0,0.2,0.6}
\definecolor{xgreen}{rgb}{0.2,0.6,0.0}
\definecolor{xred}{rgb}{0.7,0.1,0.0}
\hypersetup{colorlinks,breaklinks,citecolor=red!50, urlcolor=linkcolour, linkcolor=xgreen}

\usepackage{amsmath,amssymb,amsfonts}
\usepackage{algpseudocode}
\usepackage{xspace}
\usepackage{kotex}
\newcommand{\BfPara}[1]{{\noindent\bf#1.}\xspace}

\def\BibTeX{{\rm B\kern-.05em{\sc i\kern-.025em b}\kern-.08em
    T\kern-.1667em\lower.7ex\hbox{E}\kern-.125emX}}
\begin{document}

\title{Quantum Multi-Agent Reinforcement Learning via Variational Quantum Circuit Design}

\author{$^\dag$Won Joon Yun, $^\dag$Yunseok Kwak, $^\dag$Jae Pyoung Kim, $^\S$Hyunhee Cho, \\ $^\ddag$Soyi Jung, $^\circ$Jihong Park, and $^\dag$Joongheon Kim
\\
\IEEEauthorblockA{$^\dag$School of Electrical Engineering, Korea University, Seoul, Republic of Korea}
\IEEEauthorblockA{$^\S$School of Electronic and Electrical Engineering, Sungkyunkwan University, Suwon, Republic of Korea}
\IEEEauthorblockA{$^\ddag$School of Software, Hallym University, Chuncheon, Republic of Korea}
\IEEEauthorblockA{$^\circ$School of Information Technology, Deakin University, Geelong, Victoria, Australia}
}

\maketitle

\begin{abstract}
In recent years, quantum computing (QC) has been getting a lot of attention from industry and academia. Especially, among various QC research topics, variational quantum circuit (VQC) enables quantum deep reinforcement learning (QRL). Many studies of QRL have shown that the QRL is superior to the classical reinforcement learning (RL) methods under the constraints of the number of training parameters. This paper extends and demonstrates the QRL to quantum multi-agent RL (QMARL). However, the extension of QRL to QMARL is not straightforward due to the challenge of the noise intermediate-scale quantum (NISQ) and the non-stationary properties in classical multi-agent RL (MARL). Therefore, this paper proposes the centralized training and decentralized execution (CTDE) QMARL framework by designing novel VQCs for the framework to cope with these issues. 
To corroborate the QMARL framework, this paper conducts the QMARL demonstration in a single-hop environment where edge agents offload packets to clouds. The extensive demonstration shows that the proposed QMARL framework enhances 57.7\% of total reward than classical frameworks.   
\end{abstract}


\section{Introduction} 
The recent advances in computing hardware and deep learning algorithms have spurred the ground-breaking developments in distributed learning and multi-agent reinforcement learning (MARL)~\cite{pieee202105park}. The forthcoming innovations in quantum computing hardware and algorithms will accelerate or even revolutionize this trend~\cite{schuld2022quantum}, motivating this research on quantum MARL (QMARL). Indeed, quantum algorithms have huge potential in reducing model parameters without compromising accuracy by taking advantage of quantum entanglement~\cite{oh2020qcnn-simple}. A remarkable example is the variational quantum circuit (VQC) architecture, also known as a quantum neural network (QNN)~\cite{ijcnn21hong,icufn21kwak}, which integrates a quantum circuit into a classical deep neural network. The resultant hybrid quantum-classical model enables quantum reinforcement learning (QRL) that is on par with classical reinforcement learning with more model parameters~\cite{chen2020variational,ictc21kwak}, which can accelerate the training and inference speed while saving computing resources~\cite{carleo2019machine}. Inspired from this success, in this article we aim to extend QRL to QMARL by integrating VQC into classical MARL. This problem is non-trivial due to the trade-off between quantum errors and MARL training stability as we shall elaborate next.

In MARL, each agent interacts with other agents in a cooperative or competitive scenario. Such agent interactions often incur the non-stationary reward of each agent, hindering the MARL training convergence. A standard way to cope with this MARL non-stationarity is the centralized training and decentralized execution (CTDE) method wherein the reward is given simultaneously to all agents by concatenating their state-action pairs. In this respect, one can n\"aively implement a VQC version of CTDE as in \cite{chen2020variational}. Unfortunately, since QRL under VQC represents the state-action pairs using qubits, such a n\"ive CTDE QMARL implementation requires the qubits increasing with the number of agents, and incurs the quantum errors increasing with the number of qubits~\cite{shor1995scheme}, hindering the MARL convergence and scalability. Under the current noise intermediate-scale quantum (NISQ) era (up to a few hundreds qubits), it is difficult to correct such type of quantum errors due to the insufficient number of qubits. Instead, the quantum errors brought on by quantum gate operations can be properly controlled under NISQ~\cite{shor1995scheme}. Motivated from this, we apply a quantum state encoding method to CTDE QMARL, which reduces the dimension of the state-action pairs by making them pass through a set of quantum gates.

\BfPara{Contributions} 
The major contributions of this research can be summarized as follows.
\begin{itemize}
\item We propose novel QMARL by integrating CTDE and quantum state encoding into VQC based MARL.
\item By experiments, we demonstrate that the proposed QMARL framework achieves $57.7$\% higher total rewards compared to classical MARL baselines under a multiple edge-to-cloud queue management scenario.
\end{itemize}

\section{Quantum Computing and Circuit}\label{sec:2}
\begin{figure}[t!]
    \centering
    \includegraphics[width=1\columnwidth]{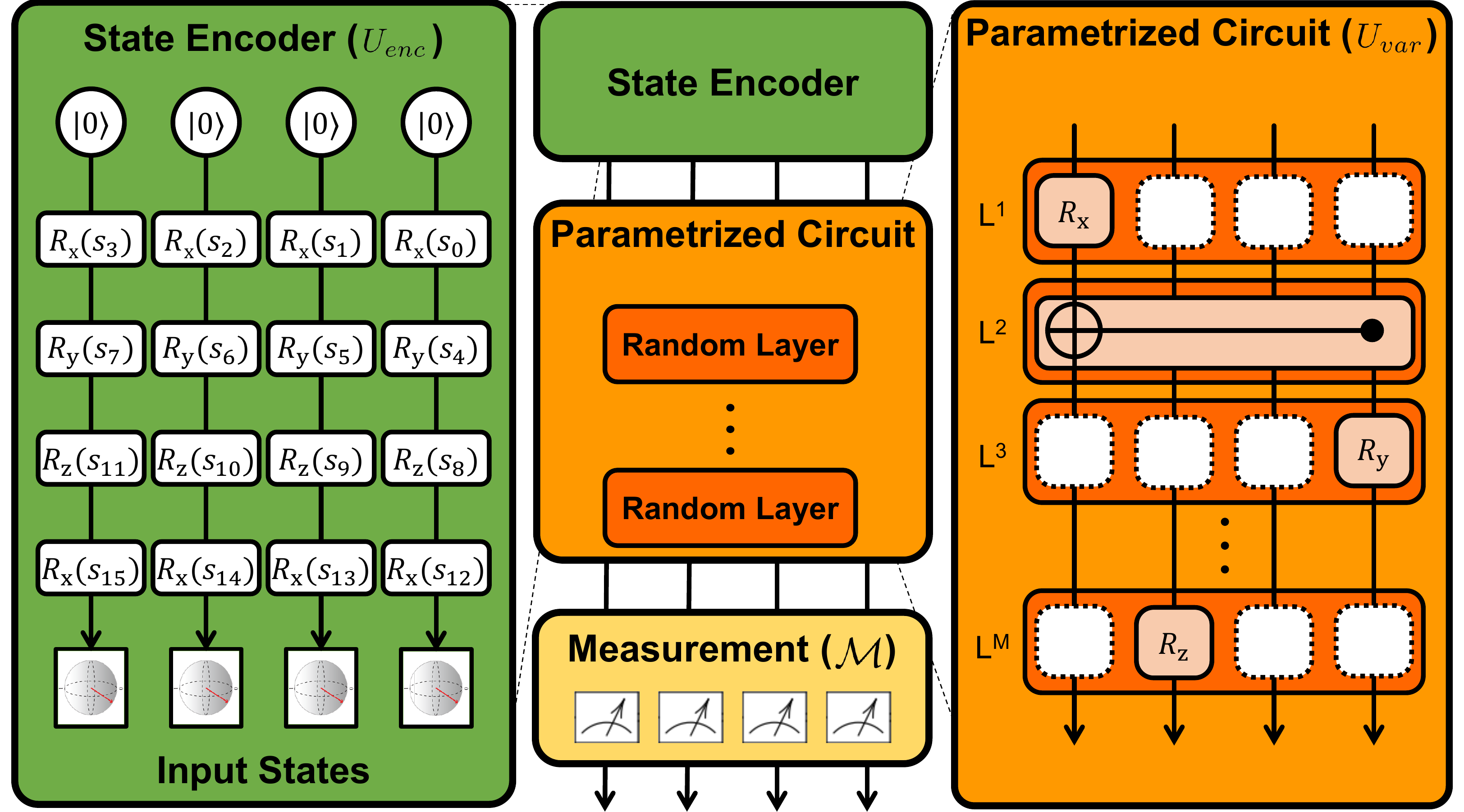}
    \caption{The illustration of VQC.}
    \label{fig:1}
    \vspace{-10pt}
\end{figure}
\subsection{Quantum Computing in a Nutshell}
Quantum computing utilizes a \textit{qubit} as the basic unit of computation. The qubit represents a quantum superposition state between two basis states, which denoted as $|0\rangle$ and $|1\rangle$. Mathematically, there are two ways to describe a qubit state:
\begin{align}
|\psi\rangle &= \alpha|0\rangle + \beta|1\rangle, ~\mathrm{where}~\|\alpha\|_2^2 + \|\beta\|_2^2 = 1\nonumber \\
|\psi\rangle &= \cos(\delta/2)|0\rangle + e^{i\varphi}\sin(\delta/2)|1\rangle, \forall \delta, \varphi \in [-\pi,\pi]. \nonumber
\end{align}

The former is based on a normalized 2D complex vector, while the latter is based on polar coordinates $(\delta,\varphi)$ from a geometric viewpoint.
The qubit state is mapped into the surface of a 3-dimensional unit sphere, which is called \textit{Bloch sphere}.
In addition, a quantum gate is a unitary operator transforming a qubit state. 
For example, $R_{\text{x}}(\delta)$, $R_{\text{y}}(\delta)$, and $R_{\text{z}}(\delta)$ are rotation operator gates by rotating $\delta$ around their corresponding axes in the Bloch sphere. These gates are dealing with a single qubit.  
In contrast, there are quantum gates which operate on multiple qubits, called controlled rotation gates. They act on a qubit according to the signal of several control qubits, which generates quantum entanglement between those qubits. Among them, a \textit{Controlled}-$X$ (or CNOT) gate is one of widely used control gates which changes the sign of the second qubit if the first qubit is $|1\rangle$. These gates allow quantum algorithms to work with their features on VQC, which will are for QMARL.

\subsection{Variational Quantum Circuit (VQC)}
VQC is a quantum circuit that utilizes learnable parameters to perform various numerical tasks, including estimation, optimization, approximation, and classification. As shown in Fig.~\ref{fig:1}, the operation of the general VQC model can be divided into three steps. The first one is \textit{state encoding step} $U_{enc}$, and in this step, a classical input information is encoded into corresponding qubit states, which can be treated in the quantum circuit. The next step is \textit{variational step} $U_{var}$, and it is for entangling qubit states by controlled gates and rotating qubits by parameterized rotation gates. This process can be repeated in multi-layers with more parameters, which enhances the performance of the circuit. The last one is \textit{measurement step} $\mathcal{M}$, which measures the expectation value of qubit state according to its corresponding computational bases. This process can be formulated as follows:
\begin{equation}
    f(x;\theta) = \otimes\Pi_{M\in\mathcal{M}}\langle0|U^{\dagger}_{enc}(x)U^{\dagger}_{var}(\theta)MU_{var}(\theta)U_{enc}(x)|0\rangle, \nonumber
\end{equation}
where $\otimes$ stands for the qubit superposition operator; $f(x;\theta)$ is the output of VQC with inputs $x$ and circuit parameter $\theta$; $\mathcal{M}$ is the set of quantum measurement bases in VQC with $|\mathcal{M}| \leq n_{qubit}$ where $n_{qubit}$ is the number of qubits.
The example of the state encoder in Fig.~\ref{fig:1} can be expressed as follows:
\begin{equation}
    U_{enc}(s_0,s_4, s_8, s_{12}) =R_\mathrm{x}(s_{12})\cdot R_\mathrm{z}(s_{8}) \cdot R_\mathrm{y}(s_{4}) \cdot R_\mathrm{x}(s_{0}).
    \nonumber
\end{equation}

The quantum circuit parameters are updated every training epoch, toward the direction of optimizing the objective output value from VQC. Through this process, VQC is known to be able to approximate any continuous function, which is similar to classical neural network computation~\cite{biamonte2021universal}. Therefore, VQC is also called \textit{quantum neural network (QNN)}~\cite{wiebe2014quantum}. In this paper, two different VQCs are used to approximate the optimal actions of actor and the accurate state value of critic.  

\section{Quantum MARL Framework}\label{sec:3}

\begin{figure}[t!]
    \centering
    \includegraphics[width=1\columnwidth]{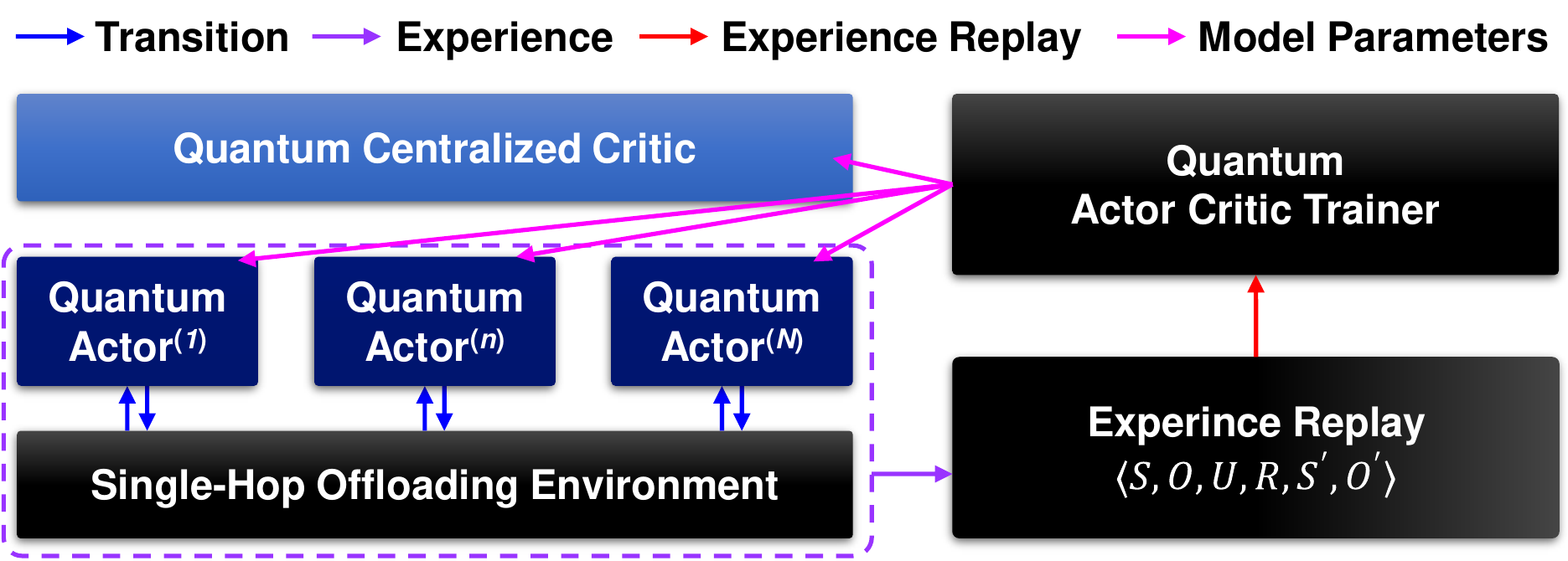}\\[-0pt]
    \includegraphics[width=1\columnwidth]{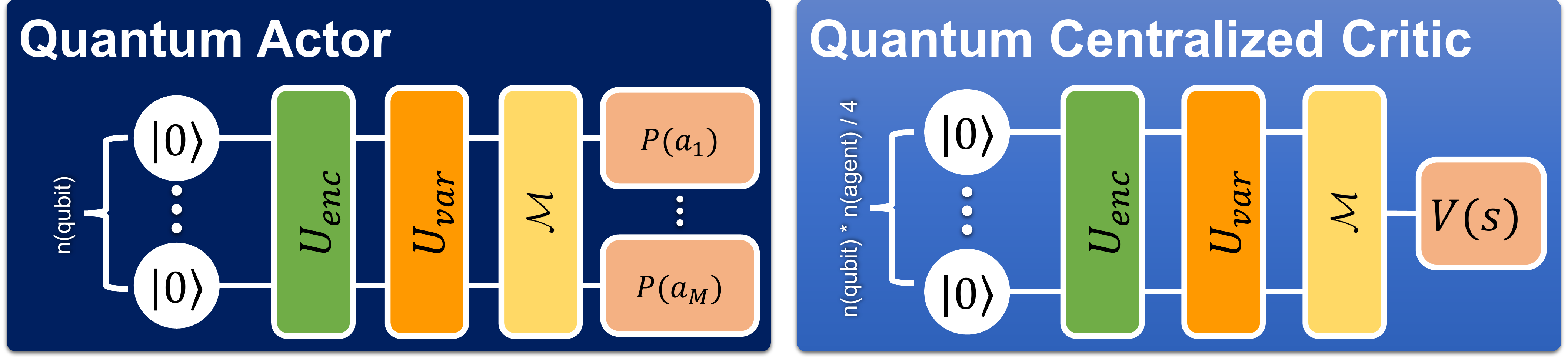}
    \caption{The structure of QMARL framework.}
    \label{fig:2}
    \vspace{-10pt}
\end{figure}

\subsection{QMARL Architecture}\label{sec:3-1} 
Our proposed QMARL is decentralized for scalability, every agent in the QMARL has a VQC-based policy, i.e., agents do not require communications among agents. Fig.~\ref{fig:2} shows the VQC that is used in quantum actor (refer to Sec.~\ref{sec:qa}) and critic (refer to Sec.~\ref{sec:qc}). 

\subsubsection{Quantum Actor}\label{sec:qa}
For the quantum actor, the VQC will be used to calculate the probabilities of actions of each agent. 
Then, the quantum policy is written as follows:
\begin{align}
    \pi_{\theta}(u_t|o_t)&=\mathsf{softmax}(f(o_t;\theta)), \nonumber
\end{align}
where $\mathsf{softmax}(\textbf{x}) \triangleq \left[\frac{e^{x_1}}{\sum_{i=1}^N e^{x_i}},\cdots,\frac{e^{x_i}}{\sum_{i=1}^N e^{x_i}},\cdots,\frac{e^{x_N}}{\sum_{i=1}^N e^{x_i}}\right]$. At time $t$, the actor policy of $n$-th agent makes action decision with the given observation $o^n_t$, which is denoted as $\pi_{\theta^n}(a^n_t|o^n_t)$. Note that $\theta^n$ denotes parameters of $n$-th actor.
Then, the action $u^n_t$ is computed as follows:
\begin{equation}
    u^n_t=\arg\max\limits_u\pi_{\theta^n}(u^n_t|o^n_t). \nonumber
\end{equation}

\subsubsection{Quantum Centralized Critic}\label{sec:qc}
We adopt the centralized critic for CTDE as a state-value function. At $t$, the parameterized critic estimates the discounted returns given $s_t$ as follows:
\begin{equation}
    V^\psi(s_t) = f(s_t;\psi) \simeq \mathbb{E}\Big[\sum_{t'=t}^{T} \gamma^{t'-t}\cdot r(s_{t'},\mathbf{u}_{t'}) \Big| s_t = s\Big],
    \nonumber
\end{equation}
where $\gamma$, $T$, $\mathbf{u}_t$, and $r(s_{t'},\mathbf{u}_{t'})$ stand for a discounted factor $\gamma \in [0,1)$, an episode length, the actions of all agents, and reward functions, respectively. In addition, $\psi$ presents trainable parameters of a critic. Note that $s_t$ is the ground truth at $t$. 
Note that the state encoding is used as shown in \textit{green box} in Fig.~\ref{fig:1} because the state size is larger than the size in observation.

\begin{algorithm}[t!]
    \footnotesize
	\caption{CTDE-based QMARL Training}
	\label{alg:1}
		Initialize the parameters of actor-critic networks and the replay buffer; {$\Theta \triangleq \{\theta^1, \cdots, \theta^N\}$, $\psi, \phi$, $\mathcal{D} = \left\{ \right\}$}\;
		\Repeat {obtaining optimal policies}{
		$t = 0, s_0=\text{initial state}$\;
    		\While{$s_t \neq terminal$ and $t < \text{episode limit}$}{
    		\For{each agent $n$}{
    		Calculate $\pi_{\theta^n}(u^n_t|o^n_t)$ and sample $u^n_t$\;
    		}
    		Get reward $r_t$ and next state and observations $s_{t+1}$, $\mathbf{o}_{t+1}= \{o^1_t,\cdots,o^N_t\}$\;
    		$\mathcal{D} = \mathcal{D} \cup \left\{(s_t, \mathbf{o}_t, \mathbf{u}_t, r_t, s_{t+1}, \mathbf{o}_{t+1} )\right\}$\;
    		{$t=t+1, \text{step}=\text{step}+1$}\;
    		}
		\For{each timestep $t$ in each episode in batch $\mathcal{D}$}{
		{Get $V^{\psi}(s_t)$; $V^\phi(s_{t+1})$}\;
		{Calculate the target $y_t$\typeout{ with \eqref{eq:target}}}\;
		}
		{Calculate $\nabla_\Theta J$, $\nabla_\psi$, 
		and update $\Theta$, $\psi$\;}
		\If{$\text{target update period}$}{
		Update the target network, $\phi \leftarrow \psi$
		}}
\end{algorithm}

\section{Experiments and Demonstrations}\label{sec:4}
\subsection{Single-Hop Offloading Environment}
The environment used in this paper consists of $K$ clouds and $N$ edges. The clouds and edges have queues $q^c$ and $q^e$ that temporally store packets. All edge agents offload their packets to clouds. 
The queue dynamics are as follows:
\begin{align}
    q^{i,k}_{t+1} &= \mathsf{clip}(q^{i,k}_{t}- u^{i,k}_{t} + b^{i,k}_{t},0,q_{\max}),\nonumber
\end{align}
where the superscript $i\in\{c,e\}$ identifies the cloud and an edge device. The terms $u^{i,k}_{t}$ and $b^{i,n}_{t}$ imply the total transmitting packet size and the packet arrival of $k$-th cloud or $n$-th edge, respectively. Note that $u^{e,n}_{t}$ is $n$-th edge agent's action. In addition, a clipping function is defined as  $\mathsf{clip}(x,x_{\min},x_{\max})\triangleq \min(x_{\max}, \max(x,x_{\min}))$.

\begin{table}[t!]
\caption{The MDP of a single-hop offloading environment.}
\centering
\begin{tabular}{l|r}
    \toprule[1pt]
    \textbf{Observation}&  $o^n_t \triangleq \{q^{e,n}_{t}, q^{e,n}_{t-1}\}\cup^K_{k=1}\{q^{c,k}_{t}\}$\\
    \midrule
    \textbf{Action}& $u^n_t \in \mathcal{A} \equiv \mathcal{I}\times \mathcal{P}$\\
    ~~$\circ$~Destination space & $\mathcal{I}\triangleq \{1,\cdots,K\}$ \\
    ~~$\circ$~Packet amount space & $\mathcal{P}\triangleq \{p_{\min}, \cdots, p_{\max}\}$ \\\midrule
    \textbf{State} & $s_t \triangleq \cup^N_{n=1}\{o^n_t\}$\\
    \midrule
    \textbf{Reward} & $r(s_t, \mathbf{u_t})$ in \eqref{eq:reward}\\\bottomrule[1pt]
\end{tabular}
\label{tab:1}
\vspace{-6pt}
\end{table}

\begin{table}[t!]
\caption{The experiment parameters.}
\centering\footnotesize
\begin{tabular}{l|r}\toprule[1pt]
    \textbf{Parameters}&  \textbf{Values}       \\\midrule
    The numbers of clouds and edge agents ($K$, $N$) & $2$, $4$ \\
    The packet amount space ($\mathcal{P}$) & $\{0.1,0.2\}$ \\
    The hyper-parameters of environment ($w_{\mathcal{P}}$, $w_\mathcal{R}$) & $(0.3, 4)$\\
    Transmitted packets from the cloud ($b^{c,k}_t$) & $0.3$\\
    The capacity of queue ($q_{\max}$) & $1$ \\
    Optimizer & Adam \\
    The number of gates in $U_{var}$  & $50$ \\
    The number of qubits of actor/critic  & $4$ \\
    Learning rate of actor/critic & $1\times 10^{-4}$, $1\times 10^{-5}$  \\\bottomrule[1pt]
\end{tabular}
\label{tab:2}
\vspace{-3mm}
\end{table}

\begin{figure}[t!]
\centering
\begin{tabular}{cc}
\multicolumn{2}{c}{\includegraphics[width=.9\columnwidth]{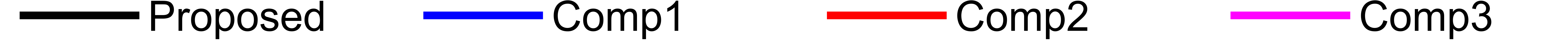}}\\[-4pt]
\includegraphics[width=.45\columnwidth]{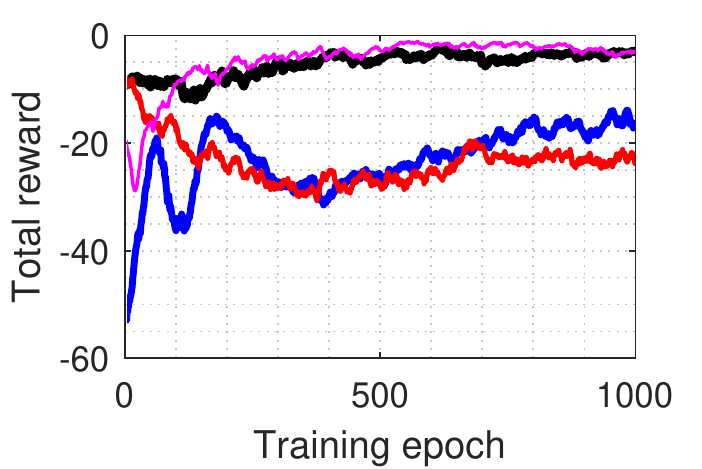} &
     \includegraphics[width=.45\columnwidth]{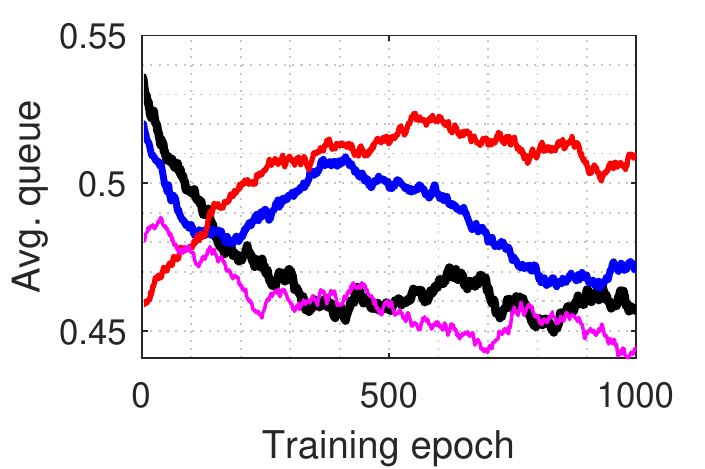} \\[-4pt]
     (a) & (b) \\[-2pt]
     \includegraphics[width=.45\columnwidth]{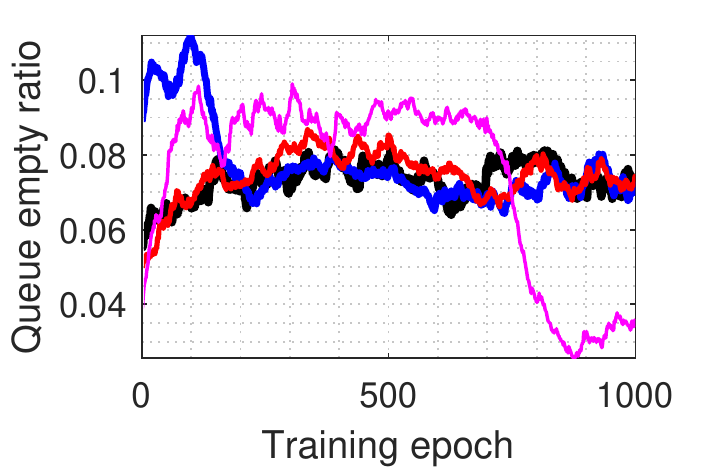} &
     \includegraphics[width=.45\columnwidth]{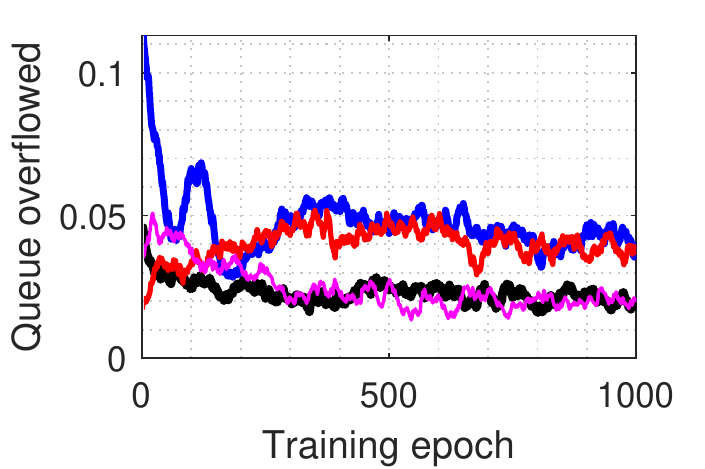} \\[-4pt]
     (c) & (d)
\end{tabular}
     \caption{The experimental result of various metrics with comparing different MARL frameworks.}
     \vspace{-3mm}
\label{fig:3}
\end{figure}
\begin{figure*}[t!]
\centering
\includegraphics[width=\linewidth]{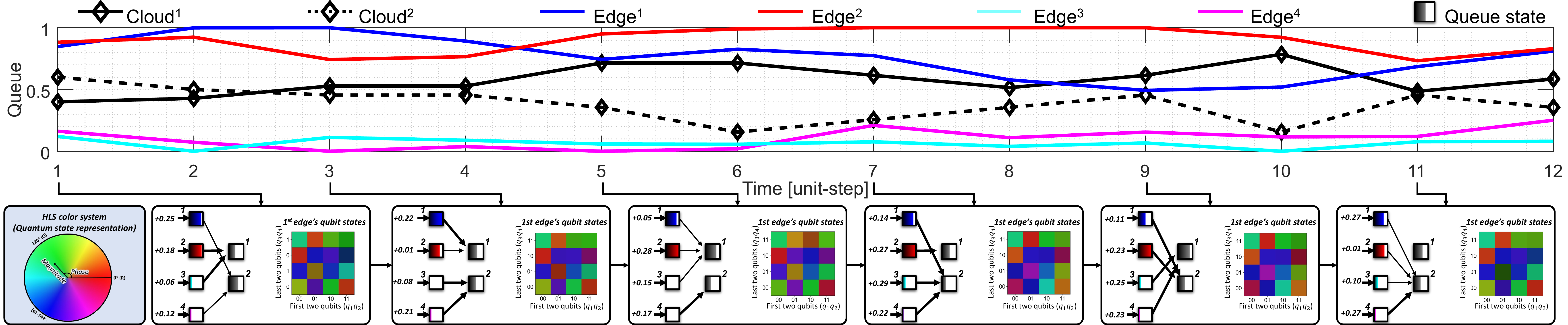}
\caption{The demonstration of QMARL framework.}
\label{fig:4}
\vspace{-5mm}
\end{figure*}

\subsection{Training}\label{sec:3-2}
The objective of MARL agents is to maximize the discounted returns. To derive the gradients, we leverage the joint state-value function $V^\psi$.
Our framework uses an multi-agent policy gradient (MAPG), which can be formulated as follows:
\begin{equation}
\nabla_{\!\theta^n} J \!=\! -\mathbb{E}_{\pi}\! \Big[ \sum\limits^{T}_{t=1}\!\sum\limits^{N}_{n=1} y_t \nabla_{\!\theta^n} \!\log\pi_{\theta}(u^n_t|o^n_t) \! \Big]\!, \nabla_\psi J \!=\! \nabla_{\!\psi}\!\sum^{T}_{t=1}\left\|y_t\right\|^2 
\nonumber
\end{equation}
s.t. $y_t =  r(s_t,\mathbf{u}_t) + \gamma V^\phi(s_{t+1}) - V^\psi(s_t)$, and $\phi$ is the parameters of target critic.
The detailed procedure is in \textbf{Algorithm~\ref{alg:1}}.

In this paper, we assume that the capacities of edges and clouds are all limited to $q_{\max}$ and edge agents receive packets from previous hops, where the distribution is uniform $\forall b^{e,n}_{t} \sim \mathcal{U}(0, w_{\mathcal{P}} \cdot  q_{\max})$. The objective of this scenario is to minimize the total amount of overflowed queue and the event that the queue is empty. Thus, the reward $r(s_t, \mathbf{u_t})$ can be as follows:
\begin{equation}
    \left.-\sum^K_{k=1} \Big[\mathbbm{1}_{(q^{c,k}_{t+1} = 0)}\cdot \tilde{q}^{c,k}_{t} + \mathbbm{1}_{(q^{c,k}_{t+1} = q_{\max})}\cdot \hat{q}^{c,k}_{t} \cdot w_\mathcal{R} \Big]\right.,
    \label{eq:reward}
\end{equation}    
s.t.
    $\tilde{q}^{c,k}_{t} = |q^{c,k}_{t}-u^{c,k}_{t}+b^{c,k}_{t}|$ and
    $\hat{q}^{c,k}_{t} = |q_{\max} - \tilde{q}^{c,k}_{t}|$, where $w_\mathcal{R}$ is the hyperparameter of rewards.
Note that $r(s_t,\mathbf{u_t}) \in [-\infty, 0]$ (negative) because we consider the occurrence of abnormal queue states (e.g., queue overflow or underflow) as a negative reward. The Markov decision process (MDP) of this environment is presented in Table~\ref{tab:1}.

\subsection{Experimental and Demonstration Setup}
To verify the effectiveness of the proposed QMARL framework (named, \textsf{Proposed}), we compare our proposed QMARL with two comparing methods. Here, `\textsf{Comp1}' is a CTDE hybrid QMARL framework where the actors use a VQC-based policy and the centralized critic uses a classical neural networks. In addition, `\textsf{Comp2}' is a CTDE classical MARL framework that is not related to quantum algorithms. Note that the trainable parameters of these three frameworks are all set to $50$ for actor and critic computation. Lastly, `\textsf{Comp3}' is a classical CTDE MARL where the number of parameters is more than 40K.
The simulation parameter settings are listed in Table~\ref{tab:2}.  We use python libraries (e.g., \texttt{torchquantum} and \texttt{pytorch}) for deploying VQCs and DL methods, which support GPU acceleration~\cite{hanruiwang2022quantumnas}. In addition, all experiments are conducted on \textsf{AMD Ryzen$^{\textsf{TM}}$ Threadripper$^{\textsf{TM}}$ 1950x} and \textsf{NVIDIA RTX 3090}. We have confirmed that the training time of QMARL for $1,000$ epochs is not expensive ($\approx$ $35$\,minutes). 
\subsection{Evaluation Results}
\subsubsection{Reward Convergence} 
Fig.~\ref{fig:3} presents the demonstration results. As shown in Fig.~\ref{fig:3}(a), the reward of QMARL frameworks is around -3.0 for \textsf{Proposed} and -16.6 for \textsf{Comp1}, whereas the classical MARL frameworks record -22.5 for \textsf{Comp2} and -2.8 for \textsf{Comp3}, respectively. 
We calculate the achievability as min-max normalization with the average returns of random walk. Note that the random walk records -33.2 on average. 
The achievability of QMARL frameworks is 90.9\% for \textsf{Proposed} and 49.8\% for \textsf{Comp1}. However, the classical MARL frameworks achieve 33.2\% for \textsf{Comp2} and 91.5\% for \textsf{Comp3}. In summary, the proposed QMARL outperforms hybrid QMARL and classical MARL under the constraint of the number of trainable parameters.
\subsubsection{Performance} The average queue states of edges/clouds and clouds are 0.460 for \textsf{Proposed}, 0.480 for \textsf{Comp1}, 0.510 for \textsf{Comp2}, and 0.453 for \textsf{Comp3}, respectively. 
The ratio of the number of empty queue events records in a high order of \textsf{Comp2}, \textsf{Comp1}, \textsf{Proposed}, and \textsf{Comp3}. 
However, the overflowed queue is low with the order of \textsf{Proposed}, \textsf{Comp3}, \textsf{Comp2}, and \textsf{Comp1}. 
According to Fig.~\ref{fig:3}(a--d), the QMARL framework outperforms both classical and hybrid quantum-classical MARL frameworks under the constraints of the number of trainable parameters.

\subsection{Demonstration}  
 Due to high network latency of utilizing quantum clouds, we conduct demonstration on simulation. Fig.~\ref{fig:4} shows the visualization of the workflow of our QMARL framework. The superpositioned qubit states (i.e., magnitude and, phase vector) are expressed as $4\times4$ heatmap in hue-lightness-saturation color system. We provide source codes\footnote{\tiny\url{https://github.com/WonJoon-Yun/Quantum-Multi-Agent-Reinforcement-Learning}} including QMARL, the single-hop environement, and the simulator.

\section{Concluding Remarks and Future Work}\label{sec:5}
This paper introduces quantum computing concepts to MARL, i.e., QMARL. To resolve the challenge of QMARL, we adopt VQC with state encoding and the concept of CTDE. From the single-hop environment, we verify the superiority of QMARL corresponding to the number of trainable parameters and the feasibility of QMARL.
As a future work direction, the implementation of QMARL to the quantum cloud (e.g., IBM quantum, Xanadu, or IonQ) should be interest because the impact of noise is considerable on quantum computing. 

\vspace{3mm}
\BfPara{Acknowledgement}
This research was supported by the National Research Foundation of Korea (2022R1A2C2004869 and 2021R1A4A1030775). W.J. Yun and Y. Kwan contributed equally to this work. S. Jung, J. Park, and J. Kim are corresponding authors. 

\bibliographystyle{IEEEtran}

\end{document}